# Changing Roles and Identities in a Teacher-Driven Professional Development Community

## Ben Van Dusen and Valerie Otero

*School of Education, University of Colorado, Boulder, 80309, USA*

**Abstract.** In a climate where teachers feel de-professionalized at the hands of regulations, testing, and politics, it is vital that teachers become empowered both in their own teaching and as agents of change. This physics education research study investigates the "Streamline to Mastery" professional development program, in which the teachers design professional development opportunities for themselves and for fellow teachers. The research reported here describes the process of teacher professional growth through changes in roles and identities. Videos, emails, and interviews were analyzed to glean insight into practice and participation shifts as these physical science teachers formed a community and engaged in their own classroom research. Implications for the role of PER in teacher professional development and teacher preparation will be discussed.

## INTRODUCTION

Our country is in dire need of highly-qualified physics teachers [1]. A majority of those who teach physics are teaching outside of their content areas, with only 35% holding a major in physics or physics education [1, 2]. Recruitment and retention is an important priority but so too is the professional development of teachers who are teaching physics outside their fields.

Streamline to Mastery is a five-year professional development program for teachers who find themselves teaching physics and physical science. The teachers themselves are in charge of establishing their own program, thus it is derived from the expertise and experiences of the teachers who participate. Too often, we undervalue, underestimate, and underutilize the vast knowledge, experience, and expertise that reside within teachers. Streamline to Mastery follows an experiential learning model where a high level of trust and a small amount of guidance is provided. This design cultivates an environment with openness, honesty, and skepticism and results in the generation of knowledge that is relevant to its participants.

The goal of Streamline to Mastery is to support teachers in improving their professional practices and to develop a community of science education leaders within the greater population of practicing teachers. These are the only formalized goals of this professional development program; more specific goals must, by design, emerge from the teachers' own perceived needs and areas of interest. This model is based on the theoretical perspective of *communities of practice* [3, 4], which defines learning in terms of participation within a community of likeminded scholars. The idea is that when a group of expert learners collaborates around a shared inquiry, knowledge is generated through skeptical discourse, individual agency, and shared experiences and prior knowledge. Within the scientific community, shared experiences include evidence, experiments, and models. In a community of teachers, shared experiences might include years of working with students, science content, administrators, and the school contexts.

In developing the Streamline to Mastery program, we hypothesized that if we could provide a context that would encourage teachers to articulate their thoughts about physics teaching and learning, challenge the ideas of other teachers, and critically analyze data from their own classrooms, they would establish a sense of agency, leadership, and enhanced knowledge regarding teaching physics. The study presented here provides preliminary data to begin to test this hypothesis. We address the following questions: (1) in what ways do Streamline teachers change their participation and leadership roles in their interactions with other teachers and researchers? and (2) in what ways does the Streamline group resemble a community of practice?

## RESEARCH CONTEXT

Four teachers from two urban high schools were recruited as the first of two cohorts of secondary physical science teachers from high needs schools to participate in Streamline. As shown in Table 1, all but one of the four teachers have been teaching three years or less and three of the four teach outside of their scientific discipline.

**TABLE 1.** Participant Demographics

| Degree | Years of Exp. | Subject Taught |
|---|---|---|
| B.A. Bio/Ph.D. Biochem | 2 | Physics |
| B.A. Bio/M.A. Urban Ed. | 3 | Physics |
| B.A. Phys/M.A. Urban Ed. | 3 | Physical Sci. |
| B.A. Bio/M.A. Science Ed. | 8 | Physical Sci. |

Requirements to be in the Streamline program include teaching in a high needs district, completion of a master's degree, and a willingness to share aspects of teaching practice and collaborate. Additionally, teachers are required to conduct research into their own practices, present at least once per year at a national education conference, and take one graduate level college course per year for credit. Teachers are given a generous stipend for participation as well as a travel allowance for conferences, and opportunities to write grant proposals to the project PI for equipment and classroom supplies. The research team consists of the NSF project PI, two doctoral students in physics education research who were formerly high school physics teachers, and one future physics teacher who is currently serving as a Noyce Fellow.

Teachers and researchers met every other week to share lessons, plan classroom research, and discuss topics of interest to the teachers. Our meetings were 2.5 to 3 hours long and were held in the evenings in teachers' classrooms. Meetings typically began with a debrief session, a chance for group members to share triumphs and difficulties since the last meeting, before moving onto the topics of the day. As part of the initial professional development process, each teacher planned an action research project that they would carry out in their classroom over the year. As the year progressed, the main topics of discussion evolved from lesson plan sharing, to designing their action research projects, to creating workshops, to planning for next year's cohort of Streamline teachers, and finally to preparing their research findings for PERC submission. Additionally, graduate and undergraduate researchers participated in a variety of ways in the teachers' classrooms, including data collection and assistance with planning and instruction.

## METHOD

In order to investigate the first research question involving the growth of teachers' participation and leadership, we coded videotapes of the meetings, analyzed email threads, and interviewed the teachers. We collected forty-two hours of video. Initially, the videotapes were coded using a generative coding scheme to determine dominant themes in the meetings. For example, we found *agenda setting* to be a useful code for analyzing teachers' changes in participation over time. In addition, email threads were analyzed to determine any increases in numbers or types of emails sent by the teachers. To address research question two, we focused on a code labeled *challenging*, which marks when a teacher or researcher challenges the statement of another group member, to evaluate both the nature of the discourse and whether it reflected what could be considered a community of practice.

In the first phase of the analysis, we examined each video to determine who had set the agendas. To do so, each segment of text in which a new agenda item was begun was coded according to whether a teacher or researcher introduced the agenda item. Agenda items were considered primary if they were related to the overarching structure of the day, such as lesson sharing, and coded secondary if they came along during conversation, for example a teacher might say, "Let's not forget to talk about the workshop." One point was given to primary agenda items and 0.5 points were given to secondary. Examples of typical agenda items include: discussing research data, planning for workshops, and planning for next year's meetings.

In order to investigate the second research question, the research team generatively developed a coding scheme to illustrate the substance of the semi-weekly meetings. The coding scheme was created through researchers individually coding the transcripts and comparing their codes to the rest of the research team. This iterative process led to creation of our complete set of codes, which have been used to inform two research projects. We arrived at twelve codes, including *challenging*, *conversation driving*, and *vulnerability*. For the purpose of this analysis, we only report on the *challenging* code. The researchers came to a consensus on all of the codes on a small subset of the transcripts; however, we are still in the process of testing for more general inter-rater reliability.

Both the first and second research questions were investigated with a second source of data, which focused on the emails sent within the Streamline group that indicated changes in participation. These emails were analyzed to determine from whom they were sent (teacher or researcher), the date of their origination, what their primary topic was, and if they represented a new line of discussion or were a reply to another email. Example discussion topics included scheduling meetings and offering resources to the group.

We triangulated our findings using a third source of data from a focus group interview that one researcher conducted with the teachers. The teachers were asked to recall how their roles and preparation for meetings had changed over time. The teachers were also asked to explain their process of preparing for two presentations they made at the Western Regional NOYCE Conference. Much of the teachers' preparation for the conference was done outside of Streamline meeting time, so their interviews gave us additional information into how they collaborated.

## FINDINGS

Analysis of emails yielded the following results. Initially, the researchers sent the majority of the emails, but over time the teachers began to send a larger share of the total number of emails. With the exception of the final month, the percentage of the emails originating from teachers increased. These results are shown in Figure 1. The total number of emails in a given month ranged from eight to fifty four. During the earlier period, designated by section (a) in Figure 1, the majority of the emails were from researchers and focused on scheduling meetings with the occasional email about meeting agenda items.

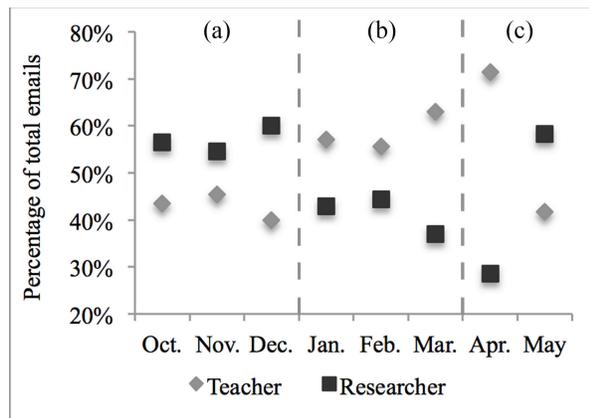

**FIGURE 1.** Percentage of total emails sent each month.

During the middle period, designated by section (b), teachers began to write more of the emails, in which they are primarily either asking for assistance or acting as a resource for one another. The March emails are dominated by the teachers' preparation for presenting a poster and holding a workshop at the Western Regional NOYCE Conference. During the late period, designated by section (c), we see teachers representing an increased percentage of the emails in April when they were sharing resources and scheduling meetings after their workshop. The month of May shows a resurgence of emails originating from researchers as they gave feedback to the teachers on their PERC papers.

We also analyzed the same set of emails to determine whether the teachers or researchers were the ones beginning the *conversation threads*. In this analysis a slightly different pattern emerges. Figure 2 shows the percentage of new email conversation threads by month, again broken down into early, middle, and late time periods. During the early time period (a) researchers began almost all of the new conversations. In the middle time period (b) conversations originating from teachers and researchers were evenly balanced with the exception of the March. Again, during March the teachers were preparing several presentations for the Western Regional NOYCE Conference, which required significant coordination among the teachers. During this time the researchers primarily acted as resources in answering teacher questions. During the late time period (c) email conversations begun by teachers and researchers balanced out.

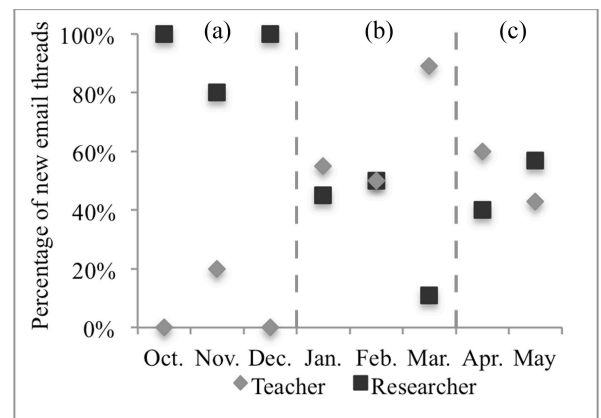

**FIGURE 2.** Percentage of new email discussion threads.

Figure 1 suggests that teacher participation increased over time and Figure 2 suggests that the teachers increased their involvement and initiative for leadership. Inferences about these findings will be made in the next section.

We now turn to findings of the video data analysis of the meetings in which we coded teachers' and researchers' contributions to meeting agenda items. To illustrate the change in agenda setting we averaged data within each time period (early, middle, and late). As shown in Figure 3(a), researchers provided all of the meeting agenda items during the early time period. The middle time period shows substantial growth in the percentage of agenda items provided by teachers. During the late section (c), the teachers provided the majority of the agenda items. The change that takes place between (a) and (b) came largely from the teachers beginning to take charge of the meetings, ultimately taking on a majority of the agenda setting responsibility.

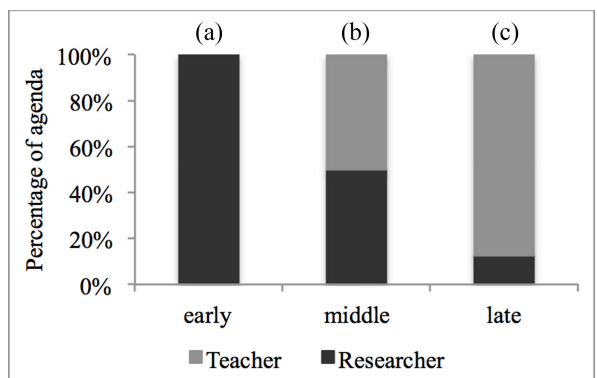

**FIGURE 3.** Agenda items from teachers and researchers.

The focus group interview confirms that teachers are aware of the changes in the community that are shown in Figures 1, 2, and 3. Teachers stated their awareness of their new roles as agenda setters, "I feel in the beginning a lot of it was kind of done for us, like we just showed up... I've noticed that Valerie (Otero) started to step out of that role of prepping the agenda and we've stepped up into that."

We turn to our investigation of the second research question involving the formation of a community of practice as defined earlier. Focusing on the challenging code in our video analysis, we found that in our early meetings neither the teachers nor researchers challenged each others' statements. Responses tended to be supportive and reassuring. During the middle time period, however, we see over a four-fold increase in the number of challenging statements. Challenging was often done explicitly, for example, T1: "Can I play devil's [advocate]?" T2: "Yeah, devil away!" By the end of the year, teachers were engaging in debates ranging from disagreements about responsibilities as Streamline teachers to whether activities were inquiry-based.

In this section of the paper we reported on four main findings of the Streamline community. First, teachers increasingly participated in the email discourse over time. Second, teachers started to begin email threads on their own. Third, teachers began to set agendas. Finally, teachers became increasingly comfortable challenging each others' ideas.

## CONCLUSION AND IMPLICATIONS

We claim that the Streamline community learned, as is evident in the changing participation of its members over time. We observed two forms of changing participation. First, an increase in total teacher participation as shown by increased emails, email thread origination, and agenda setting. Second, a shifting of roles from a hierarchical community in which the researchers were the experts and the teachers were the learners, to an egalitarian community where everyone participated equally as expert learners. In a community of practice where everyone is an expert learner, there must be constant willingness to share ideas and to challenge one another's ideas, as well as an acceptance of, and affinity for, skepticism in order for growth to take place. And indeed, this is what we observed. Over time, teachers increasingly shared their ideas and challenged one another. Our evidence supports the claim that a community of practice is forming among these teachers [3, 4] in which vulnerability and skepticism allows for the growth and development of the community as well as the growth of its individual members.

Several implications are associated with these findings. First, when we think about teacher and faculty change we might stop thinking about how to *make* people change and instead think about how to *create communities* in which change might happen. Second, when we think about professional development programs we often think about bringing expertise and resources *to* the teachers. We might instead take the view that the resources necessary for professional development reside *within* the teachers and their everyday professional experiences.

We are just scratching the surface of the data associated with this innovative program. The program suggests that as teachers develop agency they are increasingly participating in the national education discourse. As we continue to delve more deeply into this data we hope to be able to better illustrate how we might capitalize on teachers' expertise for both their own professional development and to assist in shaping educational policies.

## ACKNOWLEDGEMENTS

We are grateful to the Streamline teachers and NSF DUE #934921. We also thank Mike Ross and Sam Sherman for the work they put in to the data analysis.